\documentclass{acm_proc_article-sp}

\usepackage{url}

\begin{document}

\title{
Non-Central Limit Theorem Statistical Analysis for the "Long-tailed" Internet Society
}
%
%
%
%
%

\numberofauthors{2} 
%
\author{
%
%
\alignauthor
Kazutaka Kurihara\\
       \affaddr{National Institute of Advanced Industrial Science and Technology}\\
       \affaddr{
1-18-13 Sotokanda Chiyoda-ku}\\
       \affaddr{Tokyo 101-0021 JAPAN}\\
       \email{k-kurihra@aist.go.jp}
\alignauthor
Yohei Tutiya\\
       \affaddr{Ohara Graduate School of Accounting}\\
       \affaddr{2-2-10 Nishikanda Chiyoda-ku}\\
       \affaddr{Tokyo 101-0065 JAPAN}\\
       \email{y\_tutiya@o-hara.ac.jp}
}
\date{5 February 2010}

\maketitle
\begin{abstract}
This article presents a statistical analysis method and introduces the corresponding software package "tailstat," which is believed to be widely applicable to today's internet society. The proposed method facilitates statistical analyses with small sample sets from given populations, which render the central limit theorem inapplicable. A large-scale case study demonstrates the effectiveness of the method and provides implications for applying similar analyses to other cases.
\end{abstract}




\section{Introduction}
The spread of the modern-day internet has produced changes in traditional social structure, where large corporations and organizations supplied citizens with uniform information and materials in a one-way manner. Recent reports that more than one third of the sales of online bookseller amazon.com\cite{amazon} is made up of "unsellable books" that conventional bookstores have refused to carry, have created a stir\cite{longtail}. This is one example that counters the basic assumption that "companies increase profits by featuring popular, marketable books in the store." In addition, there are countless cases of services, such as YouTube\cite{youtube} and 
ebay auctions\cite{ebay}
, that aggregate information and materials provided by the general public. These services are also managed under completely different frameworks than the distribution/mass media industries that used to handle such business. These new structures foster and cultivate diverse content while avoiding standardization due to cost or any other ready-made ideas.  These tides and trends are expressed by the terms "Web 2.0" or "long tail," which reflect the participation of the public in the transmission of information and the increasing value of the aggregate of small-scale, diverse content. These words have become symbols of the times\cite{longtail}\cite{umeda}.

In this historical milieu, statistical methods used in social and business analysis, especially those that analyze samples from parent populations, remain limited. In conventional statistical analysis, (1) a select group of privileged analysts would (2) measure the average/overall behavior of (3) massive amounts of data. This system is typified by the opinion polls conducted by the mass media, which utilize independent information networks or financial muscle (for instance, random telephone calls to 1,000 people) to evaluate the thoughts of average 
citizens. 
Estimates of "average customer spend" used in business administration also fall into this category, using customer counts and corresponding total sales figures to measure the behavior of average customers.  These types of analysis have filled a substantial role in the large-scale, uniform, and one-way social structure outlined above.

In the long tail age, however, there are many cases to which (1), (2), and (3) do not apply.  Now, when an online bookseller sells 3 copies of an "unsellable book," let us, for instance, consider just how rare an event it is and how much market value the event has. In the present day, which places great importance on the general public's involvement in the transmission of information, this sort of statistical interest is important, but from the viewpoint of (1), "unsellable books" would not even be worth considering. In this type of analysis, contrary to (2) and (3), the sample size is too small to provide any sort of evaluation of average behavior. This is due to the fact that conventional statistical methods are founded upon the central limit theorem, which assumes a large sample size, and normal distribution sampling theory.

On the other hand, the popularization of the internet has also fueled transformations in statistical analysis. In the past, as it was difficult to obtain a population distribution for an event for analysis, research generally involved obtaining a sufficient sample and using inferential statistics for analysis. Recently, however, the benefits of information technology have made it easier to obtain population distributions, and there is growing momentum behind making the distributions public. Nico Nico Douga\cite{nikodo}, for example, releases figures for the number of video views and purchases of products related to videos, while livedoor discloses almost all of its social bookmarking service data sets\cite{ldc}. Since these examples include all of the activities in the service's economic sphere, it would be reasonable to call the sets "population distributions." 

Our research is focused on the 
rethinking
of statistical methods to make them more applicable to the age in which we now find ourselves. This article proposes an analytical method for calculating the distribution of the sample sum and sample mean when the population distribution is known. Conventionally, when the population distribution was unknown, the central limit theorem, which guarantees that the distribution of the sample sum and sample mean will have an asymptotically normal distribution, has been a powerful, useful tool. On the other hand, with a known population distribution, the sample distribution can be obtained more directly and precisely by calculating the convolution of the population distribution, regardless of the sample size. Although convolution involves a considerable amount of calculation, it can be made practical by using the tremendous capacities of recent computer technology. This is how we develop the tailstat statistical analysis software package, which conducts various analyses using the known population distribution as data to input.

It is highly likely that the use of tailstat will benefit both individual content publishers and service providers. The software allows individual content publishers to quantitatively estimate the value of their content and determine whether or not to make investments toward improving content value. Meanwhile, the software also benefits service providers by enabling data mining and information recommendation from a new perspective, one that affords a view to relatively unknown content that is likely to become more popular, as well as automatic detection of abnormal conditions, all of which contribute to a higher overall service value.

After outlining related research, this article introduces the implementation and functions of the tailstat statistical analysis software package, gives examples of analyses using tailstat, and discusses the software's effectiveness.

\section{Related Work}
In statistical analysis theory, the field that investigates known populations and the characteristics of samples obtained thereof is called descriptive statistics, while the field that estimates unknown populations based on samples is called inferential statistics\cite{SC}. As the present research examines data sets available on the internet and the statistical behavior of samples obtained from known populations, in particular the behavior of sample sums, it falls into the category of descriptive statistics. In descriptive statistics, the population distribution that individual samples follow is a given, and because each statistic is a sample variable transformation, probability distribution can, in principle, be calculated using elementary probability theory. Large sample sizes, however, due to the sheer amount of computation, make this calculation practically impossible. In such cases, one would usually obtain a corresponding asymptotic distribution for an infinite size of samples, and use it for approximation. One prominent example is the normal approximation of a sample sum or sample mean based on the central limit theorem. This kind of approximation was extremely effective in the investigation of targets characterized by (1), (2) and (3) described in the previous section because the focus was a macro analysis based on statistics obtained from a large sample size; therefore, much of the statistical work done in the 20th century onward has concentrated on asymptotic distribution. Mathematical interest probably also encouraged this trend.

However, there has not been a great deal of unified research into methods for investigating the statistical behavior of small samples, such as those treated in the present research. Exact sampling distributions for known populations are specific to each population, which researchers in the field have calculated on an ad hoc basis. This is obvious because, as stated in the previous section, only recently have descriptive statistics for small sample sizes, such as the "long tail," been recognized as valuable for having crosscutting universality. Thus there are few precedent studies that, like the present research, advocate the 
rethinking
of small sample size-specific descriptive statistics methods or the development of universal tools. It is important to point out that even statistical tools such as Microsoft Excel and SPSS are not built with the function for calculating the distribution 
of 
a sample sum for a given population distribution.

Research using mathematical models to explain the causes behind the formation of Zipfian or Pareto-type population distributions began in the 1980s. It is worth noting that the present research looks at the problem differently than this mathematical descriptive statistics model.

There have been many relevant investigations into the release of statistical data and methods to the general public in information-driven societies. Website hits and browsing history have become valuable indices in assessing site value, and methods for analyzing the behavior of site visitors are developing rapidly. One such example is Google Analytics\cite{googleanalytics}, which provides statistics and analysis for an individual site at no charge. While 
the Department of Commerce\cite{commerce}
and other governmental sites have made statistical information public for many years, new services founded on the concept of sharing various types of private-sector statistics by private initiative are beginning to appear, such as 
many eyes\cite{manyeyes}
, which 
shares 
statistical data sets and their visualizations
, and Public Data Sets on Amazon Web Services\cite{amazon_pds}, to name a few. These trends illustrate that the practice of statistics is spreading from the hands of a small number of people in the administrative realm to the general public - and that the data required for such analysis is becoming abundant.

Next we will examine several representative examples of Web 2.0 and other types of services 
to which the analysis in this article will be applied, 
organized by the relationship between the user and the service. The first example is "net shopping." Although there are some services, such as Rakuten\cite{rakuten}, that feature a collection of small, separately-run retailers on a single site, there are also services like amazon.com\cite{amazon} in which the site owner has control over the entire retail stock. In these cases, the user visits the site and pays a fee to purchase a product, but is often also able to write comments about the product's user-friendliness and other characteristics (a concept sometimes called "word-of-mouth"), a feature that is believed to affect the purchasing activity of future visitors.

Corporations are not the only ones handling products and contents, however; there are many services that collect and publish information from the general public. 
Ebay auctions\cite{ebay}
and other internet auction sites are good examples of product-centric services, but there are many sites where users are able to access content at no charge: video sharing services YouTube\cite{youtube} and Nico Nico Douga\cite{nikodo}, photo sharing site 
flickr\cite{flicker}, recipe sharing site Cookpad\cite{cookpad}, Wikipedia\cite{wikipedia}, and numerous others. Like many net shopping sites, most of these sites allow users to rate and write comments about various content. Along with total view counts, this type of user feedback is becoming a crucial element in determining the value of a given piece of content.

Another popular feature of these services is the concept of "tags:" text that can be attached to various content. Tags make it possible for content to be summarized manually and also help increase the chances for the content to appear in text-based searches. In recent years, there are separate tags and tagging systems for each service, but there are other "social bookmarking" services, such as delicious\cite{delicious} and livedoor Clip\cite{livedoorclip}, which attach tags to various URLs and improve search performance. 

Tagging and similar approaches that apply user knowledge and effort toward solving problems that would be difficult to resolve with computer systems alone are sometimes called "collective intelligence." However, in order to make the most of collective intelligence, average users must be in an environment conducive to contribution. For instance, in the tagging system described above, it is normally assumed that the content provider or viewer applies a tag consciously. However, since tagging usually requires a certain amount of effort, some services take special measures to accumulate more tags. Google Image Labeler\cite{gil}, for one, added a game-like element to its image tagging process - to the user, the experience is nothing more than a simple game, but in the background, the user's input contributes to the improvement of Google's image search. 
Another example is PodCastle, a service that allows users to perform text-based searches of podcasts.  To do this, the service collects podcasts and uses speech recognition tools to convert the audio files to text.  Then, in order to further improve the accuracy of the system, PodCastle allows users to correct errors in speech recognition.
These revisions are made
in the form of a "mistake-hunting game." Similar to the Google Image Labeler process, this added entertainment element is believed to stimulate contribution. The psychological premise that users will not be able to ignore incorrect recognition of their favorite entertainers' speech is also thought to be a driving factor behind contribution.

There are also examples that incorporate non-entertainment elements. In Q\&A sites like 
LivePerson\cite{liveperson}
, information provision is bought and sold with money or comparable service-specific currency to help fuel user participation.

We propose a statistical method that can be applied to analysis of individual content, user behavior, and by extension the characteristics of community/collective knowledge in modern-day internet services, like the above, that rely on various forms of user participation. There have been many modeled studies of social networks analyzing community growth and tagging behavior such as \cite{sbm}\cite{kdd1}\cite{kdd2}\cite{cscw1}\cite{cscw2}. For those studies we are able to provide a new descriptive statistical analysis tool. In addition, 
in the analyzed examples that follow, we define indices that evaluate contents from an unorthodox perspective, one used previously by Irie\cite{irie} and Oishi\cite{oishi08ipsj} and their teams. Irie and his team suggested "degree of edits" as a way of ranking videos on video sharing sites, while Oishi with his team evaluated social bookmark tags on the basis of "novelty," using the results to calculate the importance of the site.  Our proposal is independent of video content, tag text, and other information specific to individual sites.  Instead, it is structured to be universal, able to produce results based only on the data structures shared by various services. Thus, rather than competing with other approaches, our proposal is expected to have a synergistic effect.

\section{Statistical Package tailstat}
This section details the implementation of tailstat, a statistical analysis software package developed in C\#\cite{tailstat}.  
The package is made up of an application run from the command line and a GUI application 
, but both applications have essentially the same basic functions.

Basically, tailstat reads a probability distribution, performs convolution, and does the calculation.  Probability distributions are arranged in a csv file in the following format:
\begin{equation}
\begin{array}{c}
category_1, probability_1 \\
category_2, probability_2 \\
\vdots \\
category_k, probability_k \\
\end{array}
\end{equation}
Here, $\{category_1,category_2,\cdots\}$ is the set where the random variable takes its values, and $probability_i$ is the corresponding probability. 
This pair is stored in a hash table, with $category_i$ as the key and $probability_i$ as the value.
\begin{equation}
value \ for \ key(category_i)\ = \ probability_i
\end{equation}
A new hash table is created for the convolution of the read probability distribution, and after all values are reset to 0, convolution is achieved by obtaining 
\begin{eqnarray}
value \ for \ key(category_i + category_j)\ \nonumber\\
+= \ probability_i \times probability_j
\label{eq:convolute}
\end{eqnarray}
for all $i,j(1 \le i \le k, 1 \le j \le k)$ groups.

When using this method to perform $n$ convolutions, high $k$ or $n$ values complicate convolution due to the number of basic arithmetic operations and the fact that 
the memory required 
increases at a maximum order of $k^{n+1}$. However, as long as the distribution is not lopsided, the conditions of the central limit theorem are met before $n$ grows too high and normal distribution approximation is useful thereafter.  In addition, there are many times when $category_i$ values are spaced (for example, $0,a,2a,3a,4a,\ldots, (k-1)a$ with $a$ as a constant), in which case the total number of keys obtained after convolution is 
$(k-1)(n+1)+1$, and there is no significant memory consumption. In the future, it is thought that faster convolution calculation using FFT (Fast Fourier Transform) will be possible in such cases.

Below 
we demonstrate the 
feasibility
of tailstat through an example of binomial distribution.
In binomial distribution $Bi(n,p)$, the conventional requisites for obtaining a sufficiently accurate normal distribution approximation are $np > 5$ and $n(1 - p) > 5$. Even in lopsided cases where $p = 0.001$ or a similar value, the necessary sample number (number of convolutions - 1) $n$ is 5,000; a calculation of this size, including convolution, can be completed with tailstat on a normal notebook PC in roughly 10 seconds, and the necessary number of keys is only 5,002. Therefore, the naive convolution calculation method in equation \ref{eq:convolute} is considered sufficiently practical.

tailstat is equipped with the following functions required after convolution:
\begin{itemize}
\item Creation of a frequency distribution table based on 
any 
class interval width
\item Determination of distribution normality based on skewness and kurtosis (confirmation of the degree to which central limit theorem conditions are met)
\item Derivation of distribution right/left side probability, derivation of homogeneous probability based on normal distribution when the central limit theorem conditions are 
assumed to be
met
\item Linear transformation of probability variables
\end{itemize}
However, these functions are self-explanatory, so there is no need for individual explanation. Below is an example using tailstat in the comparison of two typical samples.

Table\ref{tab:lot} shows information regarding a Tokyo lottery drawing\cite{matsubara}. 
Let us consider the probability of person $A$, who has 1 lottery ticket, winning more than person $B$, who has 10 tickets.
$f$ is the frequency distribution given in Table \ref{tab:lot}, and $f^{*10}$ is obtained by using tailstat to perform 9 convolutions of $f$.  Here, $f^{*n}$ is defined as the 
convolutions of $n$ $f$'s. $f^{*n}$ refers to the probability distribution of the sample sum of independent $n$ samples. 
Next, "subtraction convolution" ($category_i + category_j$ changed to $category_i - category_j$ in equation \ref{eq:convolute}) gives the 
distribution $Y = X_1 - X_{10}$, where $X_1$ and $X_{10}$ denotes the winnings of A and B respectively . 
With that, the probability of $Y > 0$ is $P(Y > 0) = 0.03620$, which, when placed against a 5\% level of significance, is "rare enough to be impossible."
\footnote{
$P(Y \geq -200) = 0.7385$, so there is a very high possibility that the difference will be less than 200 yen.
}
tailstat is universal enough to easily handle such cases, where the central limit theorem does not apply and 
separate 
ad hoc
programming is required due to the difficulty of convolution calculation by hand.

\begin{table}[]
\caption{Tokyo lottery example}
\label{tab:lot}
\begin{center}
\begin{tabular}{r|r|r}
&	Winnings&Number of\\ 
&	(Japanese yen)	&winning tickets\\ \hline 
First&	40,000,000& 	7 \\
First (adjacent numbers)&	10,000,000& 	14 \\
Second&	10,000,000& 	5 \\
Third&	1,000,000& 	130 \\
Fourth&	140,000& 	130 \\
First (different group)&	200,000& 	903 \\
Second (different group)&	100,000& 	645 \\
Fifth&	10,000& 	1,300 \\
Sixth&	1,000& 	26,000 \\
Seventh&	200& 	1,300,000 \\
No Prize&	0& 	11,670,866 
\end{tabular}
\end{center}
\end{table}

\section{Analysis Examples}
This chapter provides examples of analysis using tailstat and discusses the software's effectiveness.
\subsection{ Analysis of livedoor Clip}
One practical and applicable example is the analysis of livedoor Clip, a social bookmarking service based on a system where users discover and share worthwhile sites by publicizing the bookmarks usually stored in their individual browsers. The concept of tagging is of vital importance here; as explained above, "tagging" involves assigning various relevant keywords to a bookmarked site. By using tags, users can perform tag-based searches and also recommend related sites using similar tags. Tagging is also a boon for service owners, who can use tags to improve overall service, as well as webmasters, who can use tags to boost site visibility.

Here, it is important to note that this structure has many things in common with a wide variety of internet-based services. Tagging exists not only in social bookmarking services, but also flickr, the photo sharing site, and post-/upload-based sites like Nico Nico Douga. In addition, when content quality is evaluated based on the number of corresponding comments, like the "cook report" feature on the Cookpad recipe sharing site, the number of comments has a value quite similar to the number of tags. A more direct example would be the product prices paid by the visitors to any of the myriad internet shopping sites - price is clearly an effective indicator of content value.

More generally, on these sorts of sites, one might apply a model in which the user visits content (a website), reacts to the content, and pays compensation for something. The compensation could be a tag, a comment, or a purchase price. The value of the content is determined based on the sum of the total compensation paid by visitors.

One obvious analysis using this model would be (average customer spend) = (total sales) / (number of visitors), a rudimentary theory in business management. When the number of visitors is constant, stores and products with high customer spend are considered healthy, providing the basis for many sales strategies. However, it is important to realize that these strategies are only viable when the number of visitors is at a sufficiently high level - in other words, when the conditions of the central limit theorem or the law of large numbers are met. The majority of sales in the long-tail internet shopping world (the content group) is quiet, and it is actually often unusual for the rare guest that comes along to buy at the 
same
average customer spend.

We will attempt to use tailstat to perform a significant analysis in this type of realm. The following section continues to use livedoor Clip as the example, but we would like to stress that the analysis is applicable to 
many other services that use
a "visitor/compensation" model.

\subsection{Method}
The December, 2008 livedoor Clip data set\cite{ldc} comprised user IDs, urls, "clip" time, and tag text) for 1,572,742 bookmarks. The set only includes bookmarks that existed 3 months prior to data set compilation and have been registered by 3 or more users. In our analysis, we consider bookmark registrants to be "visitors" and the number of registered tags to be "compensation." Essentially, the analysis is aimed at identifying sites that have a number of registered tags that is disproportionate (high or low) relative to the number of bookmark registrants. The discussion would be more complete if the analysis could be used to determine the percentage of site visitors that registers a bookmark (which would also be viewed as a type of compensation), but visitor figures are not included in the data set.

\begin{figure}[h]
\centering
\includegraphics[width=7cm]{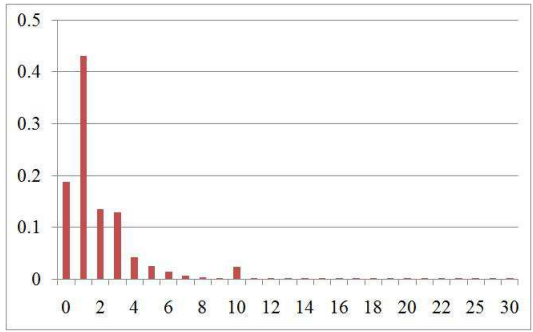} 
\caption{Frequency distribution of number of registered tags for each bookmark.}
\label{figure:master_nusers_ntag}
\end{figure} 

\begin{figure}[h]
\centering
\includegraphics[width=7cm]{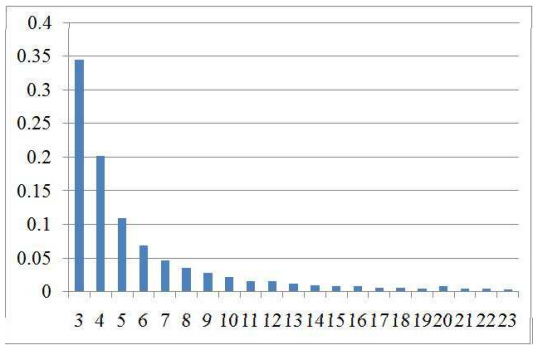} 
\caption{
Frequency distribution of bookmark registrants. Note that this visualization cuts off at 23 people but actually continues to 960 people.}
\label{figure:url_nusers_less_than_23}
\end{figure} 

Figure \ref{figure:master_nusers_ntag} is a frequency distribution that illustrates the number of tags per bookmark. Let us denote this distribution by $g$. Supposing that bookmark registration is an independent activity, the frequency distribution is basically equivalent to "the probability distribution of the compensation paid per visitor." The 1.818 expected value eventually (via the law of large numbers) becomes the average customer spend. The result is asymmetric, and there is decay after a peak at tag number 1. There is also an increase at tag number 10, which is probably because livedoor Clips allows a maximum of 10 tags and a "10 tag" culture has developed among users
\footnote{There were in fact 2,764 bookmarks with over 10 tags, but the reason for this is unknown.  We speculate that there must have been a period when livedoor allowed more than 10 tags, or there was another private method in play.}
.
Let us now move on to Figure \ref{figure:url_nusers_less_than_23}, which illustrates the main elements (23 or fewer registrants) of the frequency distribution of the number of bookmark registrants for a given site (equivalent to a "frequency distribution of the number of visitors"). Data for bookmarks with 2 or fewer registrants was not part of the data set, so it is not treated here. The distribution's expected value is 7.218 and standard deviation is 10.59, a clearly long-tailed distribution. It is thus apparent that for most sites, the number of users that will register bookmarks will be relatively low, and neither the law of large numbers nor the central limit theorem can be applied.

\begin{figure}[h]
\centering
\includegraphics[width=7cm]{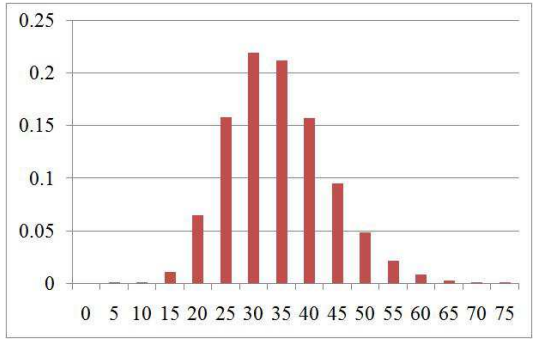} 
\caption{Histogram for $g^{*20}$ (skewness = 0.5379, kurtosis = 0.3668).}
\label{figure:liv-20}
\end{figure}

\begin{figure}[h]
\centering
\includegraphics[width=7cm]{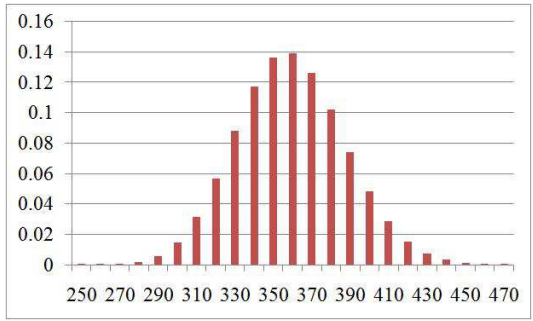} 
\caption{Histogram for $g^{*200}$ (skewness = 0.1705, kurtosis = 0.03244).}
\label{figure:liv-200}
\end{figure}

 Using tailstat, we will then obtain probability distribution $g^{*n}$, based on $n-1$ convolutions of the distribution in Figure \ref{figure:master_nusers_ntag}, relative to the number of bookmark registrants $n$ between 3 and 960, as shown in the distribution in Figure \ref{figure:url_nusers_less_than_23}. Essentially, this is the probability distribution for tag number $T$ when a bookmark has been registered $n$ times. Figures \ref{figure:liv-20} and \ref{figure:liv-200} are histograms of $g^{*20}$ and $g^{*200}$, respectively. The distribution tail is still long to the right in the $g^{*20}$ histogram, but the $g^{*200}$ distribution is almost normal. 
 Now let us choose an arbitrary bookmark with $n$ registrants and denotes its number of tags by $T$. When $T=t$ is observed, we can calculate the upper probability $p_t=P(T\ge t)$ using $g^{*n}$. 
 For comparison, we also use the normal distribution $N(n \mu, n \sigma^2), \mu = 1.818, \sigma = 2.008$  in Figure \ref{figure:master_nusers_ntag} to obtain probability $p_z=P(T\ge t)$. 
 In this case, the central limit theorem is assumed to be satisfied. With that, we will then obtain $p_t$ and $p_z$ for all of the sites in the data set. The smaller the $p_t$ and $p_z$ values, the easier it is to indicate that tags are rare; or, in other terms, that the "value is rare," or that tags are an "abnormality."
It is also possible to use $L=\prod_{j=1}^{10}p_j^{k_j}$ (where  $p_j$ denotes 
the probability that one person assigns $j$ tags and $k_j$ denotes the number of persons who assigned $j$ tags) as an index of rarity.
Many non-parametric tests based on multinomial distributions use $L$ as test statistics.Another candidate that can be readily calculated 
might be the modification $L'=\prod_{j=1}^{10}(p_j+\cdots+p_{10})^{k_j}$.
However, instead of trying to assign a particular threshold for $L$ or $L'$ to determine the rarity of the total number of tags
$\sum_{j=1}^{10}jk_j$, the tailstat approach of a probability based estimate seems more appropriate.
Another problematic point where tailstat has the advantage is that an index such as $L$ does not allow one to compare the level of rarity of a situation 
in which, for example, $3$ users assign $27$ tags among them, with a different situation where $4$ users assign $37$ tags. 


\subsection{Tailstat vs. Average Customer Spend}

\begin{figure}[h]
\centering
\includegraphics[width=7cm]{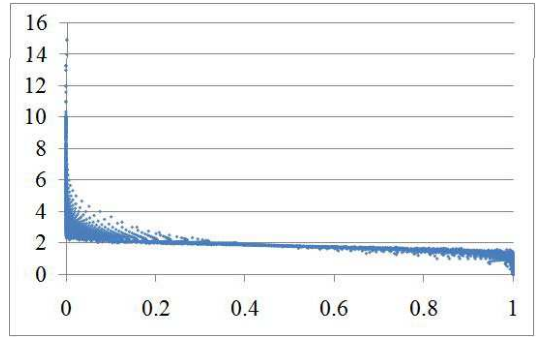} 
\caption{Average customer spend vs. $p_t$.}
\label{figure:url_xprob_uriageritsu}
\end{figure}

First, let us analyze the relationship between $p_t$ and average customer spend (total tag number $T$ / bookmark registrant number $n$), both of which may be used as indices for "sites that have a number of registered tags that is disproportionate (high or low) relative to the number of bookmark registrants."  The x-axis of Figure \ref{figure:url_xprob_uriageritsu} is $p_t$ , with average customer spend on the y-axis. The correlation can be read as basically negative, but when $p_t$ is close to zero or 1, average spend shows a large variation. What this indicates is that when a user searches for a site with a high (or low) average customer spend, the user may overlook a site with rarer value.

\begin{figure}[h]
\centering
\includegraphics[width=7cm]{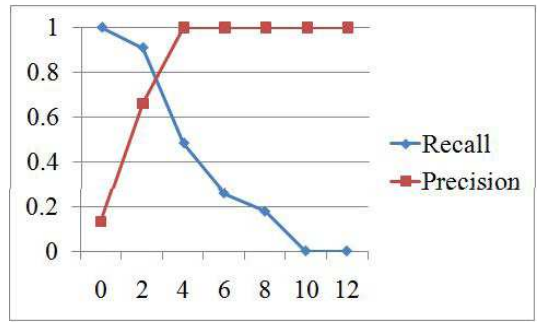} 
\caption{Precision and recall for each average customer spend as a query.}
\label{figure:uriage_recall}
\end{figure}

Figure \ref{figure:uriage_recall} is the corresponding quantitative representation, extracting only the Figure \ref{figure:url_xprob_uriageritsu} $p_t$ values less than 0.1 and defining them as "all of the sites with rare value." Figure \ref{figure:uriage_recall} illustrates how 
a "rare-valued" site can be detected  
by using average customer spend > $a$ as a query when positive parameter $a$ is increased from 0 in small increments, with a on the x-axis and recall and precision on the y-axis. When $a$ is 3, both recall and precision are slightly under 80\%, meaning that rare value sites account for roughly 80\% of the total, but sites without any rare value still accounts for over 20\%. This indicates that the average customer spend criterion cannot fully determine a site's rare value.

Also, as $p_t$ has been obtained as a probability, it has greater information volume than average customer spend. Average customer spend is often sorted in a ranking-format, using the "upper $n$-th percentile" or the "lower $n$ number of cases." This is probably because it is difficult to determine a specific, meaningful threshold along the lines of a pass/fail line, where students who score above a certain percentage pass, and those below it fail. On the other hand, $p_t$ has a specific meaning - "the probability that a similar event will occur," so it can be sorted on an ordinal scale and also from specific viewpoints, such as "rare sites that obtain tags at a probability of less than 1\%." The above results demonstrate that the $p_t$ obtained via tailstat can be used to evaluate site value in a way that average customer spend does not allow.

\subsection{tailstat vs. Central Limit Theorem}
\begin{figure}[h]
\centering
\includegraphics[width=7cm]{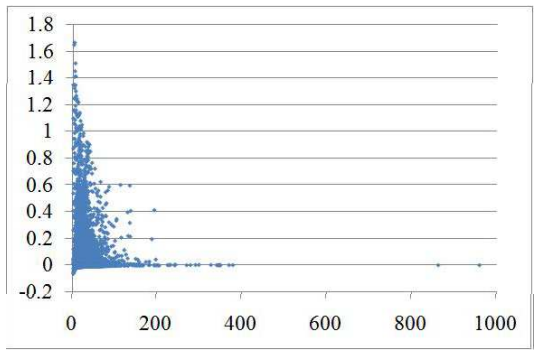} 
\caption{Number of bookmark registrants vs. $log_{10}(p_t/p_z)$.}
\label{figure:linx_liny}
\end{figure}

Next is a comparison of the resultant $p_t$ obtained via tailstat and $p_z$, calculated under the assumption that the conditions of the central limit theorem have been met. Only $p_t$ and $p_z$ values 0.0001 or higher (3,880 values) were used in order to eliminate numerical operation noise. Figure \ref{figure:linx_liny} is a scatter plot with the common logarithm of the ratio $log_{10}(p_t/p_z)$ on the x-axis and the number of bookmark registrants $n$ (sample number) on the y-axis. This figure demonstrates that when the number of bookmark registrants increases (for example, over 200), the central limit theorem becomes dominant and the difference between $p_t$ and $p_z$ grows smaller (converging to 0), but such data accounts for only 0.6\% of the total. In fact, in the majority of the plot, where the number of bookmark registrants is small, there is a substantial difference between $p_t$ and $p_z$, sometimes exceeding the single digits (values on the y-axis greater than 1). These results show that careless use of the central limit theorem should be avoided, and analysis using tailstat is very effective.

\subsection{Discussion}
Based on the above analysis, 
this section outlines the items that should be taken into consideration in generalizing our findings. This article examined the livedoor Clip tagging system as an application of a "visitor/compensation" model, but it is important to remember that this example has the following characteristics.
\begin{enumerate}
\item One visitor's compensation does not considerably affect another visitor's compensation.
\item The compensation paid by one visitor often overlaps with those of others, and this is permitted.
\item The distribution of compensation per visitor is not a normal distribution or a \textbf{\textit{reproducible}} known distribution.
\item Content that receives more compensation is able to induce more visitors.
\end{enumerate}
Below are detailed descriptions of each item.

\paragraph{Independence}
The first characteristic is that users cannot see tags registered by other users at the tagging screen.  This makes it easier to consider the bookmark registration tagging process an independent activity
\footnote{
Many users discover sites by clicking on or finding tags applied by others, so technically, this is not an independent activity.}
. On the other hand, in the case of PodCastle speech recognition error correction, it is more and more difficult to detect and correct errors as the correction process continues to move forward, meaning that compensation payment becomes less and less independent. Similarly, it is also difficult to assume independent activity in interfaces like that of Nico Nico Douga, which asks users to "add to or delete current tags."

The assumption of independence is vital to the application of tailstat or other major inferential statistics methods. When using tailstat, 
one should remember to first carefully scrutinize the subject.
On the other hand, it is also promising to deepen the analysis by naively applying tailstat and 
to be able to conclude that "the assumption of independence was inappropriate and thus these very rare results were obtained."

\paragraph{Compensation Overlap}
The second characteristic is that user tag text often overlaps with other users' tag text, and that this is permitted. This is related to the first characteristic, and a very good thing for social bookmarking services. This is because it helps 
establish correlations 
between users who apply the same tag and can lead to information recommendation and/or the development of the corresponding service. It also allows for calculation and visualization (tag clouds, etc.) that increase the importance of frequently-used tags.

The same characteristic is seen in net shopping, as well; different customers pay the same amount for an item but do not affect each other in any way.

However, in PodCastle, new visitors see what previous visitors have corrected, and overlapping corrections are not viewed as compensation. The interface does allow different users to make the same speech recognition correction, but there have been cases that show that not making identical corrections makes a better contribution to the service. With PodCastle, compensation is not viewed as "the degree of extraction of various opinions from multiple users," but "the degree of refinement of a single piece of content by multiple users."

Independence, as well as the statistical definition of the compensation for analysis, differ according to what kind of compensation the service wants from its visitors. Currently, tailstat is considered ideal for analysis of services with independence and compensation amounts that are allowed to overlap, and future research will focus on application to other types of cases.

\paragraph{Population Distribution}
The third characteristic is that the tag registration distribution (Figure \ref{figure:master_nusers_ntag}) is not a normal distribution, nor is it a well-known reproduction of the original distribution created by convolution (reproducible).  This makes tailstat more unique, as it requires numerous standard convolution calculations. If one considers a rating system that adds value to web content when a visitor gives the content a point score is to be "compensation," the population may be assumed to have a normal distribution, so tailstat is 
not necessarily required.
However, popular 5-level rating systems have the tendency of producing a ceiling or floor effect, so 
the assumption of normal distribution
is a sensitive issue.

Furthermore, 
the 
discussion in this article applies only to cases in which the population has a discrete distribution. Future research will address application to continuous distributions.

\paragraph{Analysis Feedback for Services}
The fourth characteristic is that sites with numerous tags are discovered by other users, making it more probable that more bookmarks will be added in the future. In fact, the more tags a site has, the higher the probability it will appear in searches, so other users are more likely to visit the site. This is common knowledge in the retail world - popular products are placed in easily-visible locations, and consequently sell even more. Conventional strategies arranged products (content) using indices such as sales rankings, visitor rankings, and average customer spend rankings, but it is technically simple to chronologically follow the effects that product placement based on the "rare value" index (calculated by tailstat in this article) has on subsequent sales and how it changes the rare value index, which is indeed an interesting concept. Future research will investigate these ideas.

\section{Conclusion and Outlook}
This article investigated services and data structures that have emerged in recent times due to the growing possibilities of various forms of user participation, a phenomenon often called "long tail" or "Web 2.0," and also demonstrated the importance of statistical analysis of small sample sizes with known populations to which the central limit theorem cannot be applied. The article also described 
rethinking
of sample sum/sample mean statistical methods based on the convolution of known populations, as well as the development of tailstat, a statistical analysis package that makes such restructuring possible. Then, after illustrating the effectiveness of tailstat in the analysis of livedoor Clip, the article discussed important points for consideration when applying tailstat to other cases expressed by the "visitor/compensation" model. In the future, we plan to explore analysis methods for items besides sample sum/sample mean, expand tailstat, improve calculation speed, and add functions for continuous distribution analysis. We also hope to accumulate further analysis examples through the release of the application\cite{tailstat}.


\section{Acknowledgments}
We would like to thank livedoor for providing valuable data.

%
\bibliographystyle{abbrv}
\bibliography{sigproc-sp}  
%
%
\balancecolumns
\end{document}